\documentclass[12pt]{article}

\usepackage{textcomp}
\usepackage{amsfonts}
\usepackage{amssymb}
\usepackage{verbatim}

\usepackage{mathrsfs}
\usepackage{multirow}

\usepackage{epsfig}
\usepackage{amsmath}
\usepackage{mathtools}
\usepackage{multirow}
\usepackage{amsthm}
\usepackage{lscape}
\usepackage{natbib}
\usepackage{graphicx}

\usepackage{hhline}

\newtheorem{theorem}{Theorem}
\newtheorem{corollary}[theorem]{Corollary}

\begin{document}


\title{Estimating restricted mean treatment effects with stacked survival models}

\author{Andrew Wey \\University of Hawaii \and David M. Vock, John Connett, and Kyle Rudser \\ University of Minnesota}

\date{ }



\maketitle

\begin{abstract}
{The difference in restricted mean survival times between two groups is a clinically relevant summary measure.  With observational data, there may be imbalances in confounding variables between the two groups.  One approach to account for such imbalances is to estimate a covariate-adjusted restricted mean difference by modeling the covariate-adjusted survival distribution and then marginalizing over the covariate distribution.  We demonstrate that the mean-squared error of the restricted mean difference is bounded by the mean-squared error of the covariate-adjusted survival distribution estimators.  This implies that a better estimator of the covariate-adjusted survival distributions is associated with a better estimator of the restricted mean difference.  Thus, this paper proposes estimating restricted mean differences with stacked survival models.  Stacked survival models estimate a weighted average of several survival models by minimizing predicted error.  By including a range of parametric and semi-parametric models, stacked survival models can effectively estimate a covariate-adjusted survival distribution and, therefore, the restricted mean treatment effect in a wide range of scenarios.  We demonstrate through a simulation study that the new estimator can perform nearly as well as Cox regression when the proportional hazards assumption is satisfied and significantly better when proportional hazards is violated.  The proposed estimator is also illustrated with data from the United Network for Organ Sharing to evaluate post-lung transplant survival between large and small-volume centers.}

\noindent
{\bf Keywords}: Bias-variance tradeoff; proportional hazards assumption; restricted mean difference; stacked survival models; survival analysis.
\end{abstract}



\section{Introduction}
Patients with end-stage lung disease (e.g., advanced cystic fibrosis) may be eligible for lung transplantation after other treatment options fail.  Unfortunately, post-transplant survival is poor after lung transplantation, especially in comparison to other solid organ transplants, with one and three-year graft survival of $79\%$ and $64\%$, respectively.  Given the poor prognosis, understanding the factors related to post-transplant survival remains an important but controversial task.  For example, previous research has suggested that transplant center volume (which we operationalize as the number of lung transplants performed at a center over the preceding two years) is associated with post-transplant survival with higher volume centers associated with lower mortality \citep{weiss:others:2009, thabut:others:2010}.  However, the effect of center volume after properly adjusting for confounders remains unclear.

In the absence of censoring, the difference in post-transplant survival time between high and low volume centers would be traditionally summarized by the difference in mean survival time.  However, the mean for a non-negative random variable (e.g., survival time) is defined as $E\{ T \} = \int_0^{\infty} S(t )dt$, where $S(t ) = P(T > t )$ is the survival function of the random variable $T$.  The estimate of the mean is, therefore, not defined when $\hat{S}(t ) > 0$ for all observed $t$, a situation regularly experienced with even light censoring.  Furthermore, there exists no consistent estimator of $E\{ T \}$ when $S(t_{max})$, where $t_{max}$ is the maximum follow-up time.  Since substantial censoring is experienced in lung transplantation, a different summary measure is required.

The restricted mean is an alternative summary measure that is always estimable in the presence of right-censored survival times \citep{royston:parmar:2013}.  The $\tau$-restricted mean truncates observations at some time point $\tau$ [i.e., $E\{ \min(T, \tau) \} = \int_0^{\tau} S(t )dt$].  By choosing a value for $\tau$ within the observed follow-up time, the restricted mean is an estimable summary measure with a direct interpretation that is closely related to the mean.  For example, the average difference in post-transplant survival over one year between high-volume and low-volume centers is the difference in one-year restricted means between the two groups.  However, estimating the difference of the restricted mean survival time with data from observational studies, such as the data available in the lung transplantation example, is difficult due to potential confounding.  In particular, the difference in the area under the Kaplan-Meier survival curve up to time $\tau$ is not necessarily a consistent estimator of the causal restricted mean difference between the two treatment groups.

To account for imbalances in confounders between the two treatments, several researchers have proposed estimating the covariate-adjusted restricted mean difference by modeling the covariate-adjusted survival distribution (i.e., estimating the conditional survival function), and then marginalizing over the covariate distribution to estimate restricted mean difference (referred to as the ``regression'' approach).  For such an approach, the model of the covariate-adjusted survival distribution (or the estimator of the conditional survival function) defines the estimator of the restricted mean difference.  


The covariate-adjusted survival distribution is often assumed to follow a proportional hazards model.  For example, \citet{karrison:1987} proposed modeling the survival time distribution as a proportional hazards model with a piecewise constant baseline hazard function.  As a natural alternative, \citet{zucker:1998} proposed a proportional hazards model with an unspecified baseline hazard, i.e., a Cox proportional hazards model~\citep{cox:1972}.  Both Karrison and Zucker assume that the proportional effect of the covariates on the hazard is the same for each treatment.  \citet{chen:tsiatis:2001} relax this assumption by estimating separate baseline hazard functions and covariate effects for each treatment.

Unfortunately, the proportional hazards assumption may not hold in many applications.  For example, centers with greater lung transplant volume are more likely to perform bilateral lung transplants (as opposed to single lung transplants), and the type of lung transplant is well-known to violate the proportional hazards assumption \citep{thabut:others:2009}.  As such, current approaches, which rely on the proportional hazards assumption, may produce biased and inefficient estimates of the restricted mean difference between high volume and low volume centers.  As an alternative, we could estimate the survival distribution with a different model (e.g., an accelerated failure time model), but the estimator would then be biased if the accelerated failure time assumption is violated.  Rather than rely on a particular parametric or semi-parametric model, we pursue a flexible estimator of the conditional survival function and, therefore, of the restricted mean difference that performs well across a wide range of situations.

In particular, we investigate `stacked survival models' for estimating the conditional survival function.  Stacking finds a weighted average of several conditional survival function estimators by minimizing predicted error \citep{wey:others:2013}.  Since the minimization is based on predicted error, stacking can include parametric models, semi-parametric models and non-parametric models.  This allows more weight to be given to the model that most accurately estimates the underlying survival function for a given situation and sample size.  \citet{wey:others:2013} demonstrated that the stacked survival model is competitive with parametric models and the Cox proportional hazards model for estimating the conditional survival function when assumptions are satisfied, but performs better when assumptions are violated.  We show that the mean-squared error of the restricted mean difference estimator is bounded by the mean-squared error of estimator the conditional survival function.  As such, better estimators of the conditional survival function will generally lead to a better estimator of the restricted mean difference.  Therefore, our goal is the improvement of restricted mean treatment effect estimation in a wide range of situations by estimating the conditional survival function with stacked survival models.

Section \ref{rm:estimator:sec} introduces the estimator of the restricted mean treatment effect and bounds the mean-squared error (MSE) of the restricted mean treatment effect by the mean-squared error of the conditional survival function.  Section~\ref{stacking:sec} outlines how stacked survival models can be applied to restricted mean treatment effect estimation.  A simulation study evaluates the finite sample performance of the proposed estimator in Section \ref{rm:sim:study}.  The proposed estimator is then applied to a observational registry of post-lung transplantation survival from the United Network for Organ Sharing in Section \ref{rm:LT:applied}.  Concluding remarks are presented in Section~\ref{rm:conclusion:sec}.

\section{Proposed Estimator}
\label{rm:estimator:sec}
Throughout the paper, random variables and observed variables are distinguished by capital and lower case letters, respectively.  The factor (i.e., the treatment or condition), which the restricted mean survival is compared across, is denoted by $a_i$, where $i$ denotes the subject, and follows the Bernoulli random variable $A$ (i.e., $A = \{ 0, 1\}$).  Additional covariates, denoted by vector $\boldsymbol{x}_{i}$, are measured at the beginning of the study and follow the distribution of the random variable $\boldsymbol{X}$.  For this paper, we define the non-negative survival time random variable as $T = T^0 \cdot I(A = 0) + T^1 \cdot I(A = 1)$, where $T^0$ and $T^1$ are the (possibly unobservable) survival time random variables had a patient received treatment $0$ and $1$, respectively.  We assume that there are no unmeasured confounders; that is, the set of potential outcomes, $(T^0, T^1)$, is conditionally independent of $A$ given $\boldsymbol{X}$ (i.e., $(T^0,T^1) \bot~A~|~\boldsymbol{X}$, where $\bot$ denotes statistical independence).  The censoring time is $c_i$, which follows the distribution of the continuous non-negative random variable $C$ and is assumed to be conditionally independent of $(T^0, T^1)$ (i.e., $(T^0,T^1)~\bot~C~|~\{\boldsymbol{X}, A\}$).    Hence a sample of right censored survival data for $n$ patients is $\{y_{i}, \delta_{i}, a_i, \boldsymbol{x}_{i} \}$, $i = 1,...,n$, where $y_{i} = \text{min}(t_{i},c_{i})$ and $\delta_{i} = I(t_{i} < c_{i})$.  

Now let $S^{(A = a)}(t | \boldsymbol{X} = \boldsymbol{x}) = P(T > t | \boldsymbol{X} = \boldsymbol{x}, A = a)$ and $G^{(A = a)}(t | \boldsymbol{X} = \boldsymbol{x}) = P(C > t | \boldsymbol{X} = \boldsymbol{x}, A = a)$ be, respectively, the treatment-specific conditional survival functions for the covariate-adjusted survival and censoring distributions for treatment $a$.  For brevity, we write throughout that $S^{(A = a)}(t | \boldsymbol{X} = \boldsymbol{x}) = S^{(a)}(t |\boldsymbol{x})$ and $G^{(A = a)}(t | \boldsymbol{X} = \boldsymbol{x}) = G^{(a)}(t |\boldsymbol{x})$.

\subsection{Restricted Mean Treatment Effects}
\label{sub:sec:rmte}
Following the outline of \citet{chen:tsiatis:2001}, we estimate the restricted mean for each treatment group with the ``regression'' approach, which involves modeling the covariate-adjusted survival time distribution.  In particular, the restricted mean for treatment $a$ is defined as
\begin{eqnarray}
\label{rm:def}
\mu(\tau, a) \equiv E\{ \min (T^a, \tau ) \} &=&E_{\boldsymbol{X}} [ E\{ \min (T^a, \tau ) | \boldsymbol{X} = \boldsymbol{x} \} ]  \nonumber \\
&=& E_{\boldsymbol{X}} [ E\{ \min (T, \tau ) | \boldsymbol{X} = \boldsymbol{x}, A = a \} ]  \\
&=& E_{\boldsymbol{X}} \left \{ \int_0^{\tau} S^{(a)}(t | \boldsymbol{x}) dt  \right \},  \nonumber
\end{eqnarray}
where (\ref{rm:def}) holds due to the assumption that $(T^0,T^1) \bot A | \boldsymbol{X}$ (i.e., the assumption of no unmeasured confounders).  It is important to note that the outer expectation in (\ref{rm:def}) is taken with respect to the marginal, rather than conditional, covariate distribution.  

Once we estimate the treatment-specific conditional survival functions, $S^{(a)}(t | \boldsymbol{x})$, then we estimate the expectation over the covariate space with the empirical covariate distribution.  Thus, the estimator for the $\tau$-restricted mean of the potential outcome $T^a$ is
\begin{eqnarray}
\label{rm:estimate}
\hat{\mu}(\tau, a) &=& \frac{1}{n} \sum_{i = 1}^n \int_0^{\tau} \hat{S}^{(a)}(t | \boldsymbol{x}_i) dt,
\end{eqnarray}
where $\hat{S}^{(a)}(t | \boldsymbol{x}_i)$ is the estimate of $S^{(a)}(t | \boldsymbol{x}_i)$.  In practice, a closed form solution of equation (\ref{rm:estimate}) may not exist, and we, therefore, approximate equation (\ref{rm:estimate}) by
\begin{eqnarray}
\label{rm:trt:l}
\hat{\mu}(\tau, a) &\approx& \frac{1}{n} \sum_{i = 1}^n \sum_{j = 1}^{N_\tau} (t_{(j)} - t_{(j - 1)}) \times \hat{S}^{(a)}(t_{(j-1)} |  \boldsymbol{x}_i),
\end{eqnarray}
where $t_{(j)}$ are the ordered event times, i.e., $t_{(j)} - t_{(j-1)} \geq 0$ for all $j = 1..., n$, and $N_\tau$ is one more than the number of event times less than $\tau$.  If no event time equals $\tau$ (i.e., $t_{(j)} \not= \tau$ for all $j = 1,...,n$), then $t_{(N_\tau)} = \tau$ and $t_{(N_\tau - 1)}$ is the largest event time less than $\tau$.  For two treatments, the estimated difference in restricted mean survival time~is
\begin{eqnarray}
\label{rm:trt:effect}
\hat{\gamma}(\tau) &=& \hat{\mu}(\tau, a = 1) - \hat{\mu}(\tau, a = 0),
\end{eqnarray}
which also corresponds to the difference in the area under the estimated survival curves for the two potential outcomes up to time $\tau$.

\subsection{Influence of Treatment-Specific Conditional Survival Functions}
\label{cdf:influence}
There is a clear connection between the restricted mean treatment effect and the treatment-specific conditional survival functions.  We further formalize this connection by placing an upper bound on the MSE of the restricted mean treatment effect estimator that depends on the MSE of the estimators for the treatment-specific conditional survival functions.
\begin{theorem}
\label{rm:inequality}
Define the mean-squared error of a restricted mean treatment effect estimator as $\text{MSE}[\hat{\gamma}(\tau)] = E [\hat{\gamma}(\tau) - \gamma(\tau)]^2$, then 
\begin{eqnarray}
\text{MSE}[\hat{\gamma}(\tau)] &\leq& \tau \times \left[ \text{MSE}_{\tau} \{ \hat{S}^{(0)}( \cdot | \boldsymbol{x}) \} + \text{MSE}_{\tau} \{ \hat{S}^{(1)}( \cdot | \boldsymbol{x}) \} \right], \nonumber
\end{eqnarray}
where $\text{MSE}_{\tau} \{ \hat{S}^{(a)}( \cdot | \boldsymbol{x}) \} = E \int_0^{\tau} [ \hat{S}^{(a)}(t | \boldsymbol{x}) - S^{(a)}(t | \boldsymbol{x}) ]^2 dt$ is the mean-squared error of an estimator of the treatment-specific conditional survival function for treatment $a$.  Note that the expectation is over the random variable for the covariate space and the sampling distribution of the conditional survival function estimator.
\end{theorem}
The inequality (see the Supporting Information for the proof) illustrates that the MSE of the restricted mean treatment effect should be associated with the MSE of the treatment-specific conditional survival functions.  Of course, mean-squared error decomposes into a squared bias term and a variance term, which implies the following corollary
\begin{corollary}
Define $\text{Bias}[\hat{\gamma}(\tau)] = E\{ \hat{\gamma}(\tau) - \gamma(\tau) \}$ and $\text{Var}[\hat{\gamma}(\tau)] = E \{ \hat{\gamma}(\tau) - E\hat{\gamma}(\tau) \} ^ 2$ as, respectively, the bias and variance of the restricted mean treatment effect, then 
\begin{eqnarray}
\text{Bias}[\hat{\gamma}(\tau)]^2 &\leq& \tau \times \left[ \text{MSE}_{\tau} \{ \hat{S}^{(0)}( \cdot | \boldsymbol{x}) \} + \text{MSE}_{\tau} \{ \hat{S}^{(1)}( \cdot | \boldsymbol{x}) \} \right] - \text{Var}[\hat{\gamma}(\tau)] \nonumber \\
&<& \tau \times \left[ \text{MSE}_{\tau} \{ \hat{S}^{(0)}( \cdot | \boldsymbol{x}) \} + \text{MSE}_{\tau} \{ \hat{S}^{(1)}( \cdot | \boldsymbol{x}) \} \right],  \nonumber
\end{eqnarray}
since $\text{Var}[\hat{\gamma}(\tau)] > 0$.
\end{corollary}
Thus, the performance of the treatment-specific conditional survival functions places an upper bound on both traditional measures of performance (i.e., bias and MSE) for the restricted mean treatment effect.  The bound on the squared bias, or absolute bias, is less tight than the bound on the MSE due to a positive, and potentially large, variance term.  Therefore, we would expect a relatively strong association between the MSE of the restricted mean treatment effect and the MSE of the treatment-specific conditional survival functions, but a relatively weak association between the bias of the restricted mean treatment effect and the MSE of the treatment-specific conditional survival functions.  The simulation study in Section \ref{rm:sim:study} presents an empirical demonstration of this relationship.

\section{Stacked Survival Models}
\label{stacking:sec}
Since the performance of the restricted mean treatment effect estimator is closely related to the performance of the treatment-specific conditional survival function estimator, we propose estimating restricted mean treatment effects with stacked survival models.  Stacked survival models estimate a weighted combination of survival models that can span parametric, semi-parametric, and non-parametric models by minimizing prediction error.  For estimating conditional survival functions, non-parametric estimators can be preferred to parametric and semi-parametric estimators due to relaxed assumptions.  Yet, even when misspecified, parametric and semi-parametric estimators can possess better operating characteristics in small sample sizes due to smaller variance than non-parametric estimators.  Fundamentally, this is a bias-variance tradeoff situation and, by minimizing predicted error, stacking estimates an optimal combination of survival models that balances the bias-variance trade-off of each estimator at a given sample size.  In particular, \citet{wey:others:2013} illustrate that stacked survival models effectively estimate a conditional survival function across a wide range of situations.

In uncensored settings, stacking estimates the optimal weighted average of several candidate models by minimizing predicted squared error \citep{breiman:stacking:1996}.  However, predicted squared error is often poorly defined in censored settings due to potentially unobserved event times. Therefore, a different loss function that is tailored to censored data is required.  \citet{wey:others:2013} evaluate the Brier Score \citep{graf:others:1999} over a grid of time points for stacking survival models.  In addition, they show that, under certain conditions, the stacked survival model using the Brier Score is uniformly consistent for the true conditional survival function.  This section focuses on appropriately modifying stacked survival models for estimating restricted means.

The Brier Score \citep{graf:others:1999} measures the predicted squared error of a conditional survival function at a particular time point.  Following \citet{lostritto:strawderman:molinaro:2012}, the estimated Brier Score for a given estimator of the conditional survival function for treatment $a$ at a single time point $t$ can be written as
\begin{eqnarray}
\label{brier:score:t}
\widehat{BS}^{(a)}(t) &=&  \frac{1}{n} \sum_{i \epsilon \Gamma_a} \frac{\Delta_i(t)}{\hat{G}^{(a)}(\min\{t_i, t\} | \boldsymbol{x}_i)} \times \{ Z_i(t) - \hat{S}^{(a)}(t | \boldsymbol{x}_i) \}^2,
\end{eqnarray}
where $Z_i(t) = I(t_i > t)$, $\Delta_i(t) = I(\min \{t_i, t\} \leq c_i)$, $\hat{G}^{(a)}(\cdot | \boldsymbol{x}_i)$ is the estimated conditional survival function of the censoring distribution for subjects that received the $a^{th}$ treatment, and $\Gamma_a$ is the set of patients that received treatment $a$.  For a fixed time $t$, censored observations with $c_i > t$ will contribute to the Brier Score, but when $c_i < t$ the censored observations will only contribute to the Brier Score indirectly through the estimation of $G^{(a)}(T_i(t) | \boldsymbol{x}_i)$.  In this paper, $\hat{G}^{(a)}(\cdot | \boldsymbol{x}_i)$ is the (unconditional) treatment-specific Kaplan-Meier survival estimator denoted here after as $\hat{G}^{(a)}(\cdot)$.

To estimate the conditional survival function for each treatment group, the stacking procedure considers the same set of $m$ candidate models, and each model has a corresponding conditional survival function estimate, say $\hat{S}_k^{(a)}(t | \boldsymbol{x}) \text{ for } k = 1,...,m$ survival models.  The set of candidate survival models included in the stack has an influence on the performance of the stacked estimator.  In particular, \citet{breiman:stacking:1996} and \citet{wey:others:2013} found that stacking performs well with a diverse set of models.  We note, however, that other combinations of survival models could be included in the stack.  Since the goal is estimating the entire conditional survival function up to time $\tau$, stacked survival models minimize the sum of $\widehat{BS}^{(a)}(t)$ over a set of time points, say $\{t_1,...,t_s \}$.  The stacking weights for the $m$ models are estimated by a weighted least squares problem with the additional constraints that $\hat{\alpha}^{(a)}_k \ge 0$ for $k = 1,..., m$ and $\sum_{k = 1}^m \hat{\alpha}^{(a)}_k = 1$:
\begin{eqnarray}
\label{bs:stacking:t}
\hat{\boldsymbol{\alpha}}^{(a)} &=& \text{arg} \min_{\boldsymbol{\alpha}^{(a)}, \alpha^{(a)}_k \ge 0} \sum_{r = 1}^{s} \sum_{i \epsilon \Gamma_a} \frac{\Delta_i(t_r)}{\hat{G}^{(a)}(\min\{t_i, t_r \})} \times \{ Z_i(t_r) - \sum_{k = 1}^{m} \alpha_k \hat{S}_k^{(a,-i)}(t_r | \boldsymbol{x}_i) \}^2,
\end{eqnarray}
where $\hat{S}_k^{(a,-i)}(t | \boldsymbol{x}_i)$ is the survival estimate from the $k^{th}$ model while leaving the $i^{th}$ observation out during the fitting process.  This ensures that stacking does not reward models that over-fit the data.  This is traditionally done by leaving only the $i^{th}$ observation out in the fitting process.  However, the computational requirements induced by bootstrapping for confidence interval estimation (which is discussed subsequently) prevent fitting the set of candidate survival models $n$ separate times.  The data are instead randomly split into five equally sized sets and $\hat{S}_k^{(a,-i)}(t | \boldsymbol{x}_i)$ is obtained for observations in a given set by fitting the survival models to the observations in the other four sets.  As such, five sets of survival models, rather than $n$ sets of survival models, are fit for each of the $m$ candidate survival models.  

After minimizing equation (\ref{bs:stacking:t}), the stacked estimate of the conditional survival function for treatment $a$ is
\begin{eqnarray}
\label{stacked:surv:est}
\hat{S}^{(a)}(t|\boldsymbol{x}) &=& \sum_{k = 1}^{m} \hat{\alpha}_k^{(a)} \hat{S}_k^{(a)}(t | \boldsymbol{x}),
\end{eqnarray}
where $\hat{S}_k^{(a)}(t | \boldsymbol{x})$ is the $k^{th}$ survival model estimated with all observations on treatment $a$.  The treatment-specific restricted means and the restricted mean treatment effects are then estimated with equations (\ref{rm:trt:l}) and (\ref{rm:trt:effect}), respectively.

We estimate confidence intervals with the non-parametric bootstrap \citep{efron:tibs:1993}.  In particular, we randomly sample with replacement $n$ of the observed $\{y_{i}, \delta_{i}, a_i, \boldsymbol{x}_{i} \}$; this is called the $b^{th}$ bootstrap data set.  The $b^{th}$ bootstrapped estimate of the $k^{th}$ treatment-specific conditional survival function is defined as $\hat{S}_{k,b}^{(a)}(t|\boldsymbol{x})$.  Since the stacking weights are re-estimated for each bootstrap, the final estimate of the conditional survival function for treatment $a$ with data from the $b^{th}$ bootstrap is $\hat{S}^{(a)}_b(t|\boldsymbol{x}) = \sum_{k = 1}^{m} \hat{\alpha}_{k,b}^{(a)} \hat{S}_{k,b}^{(a)}(t | \boldsymbol{x})$.  The $b^{th}$ bootstrap estimates for the treatment-specific restricted means and restricted mean treatment effect are calculated as before [see equations (\ref{rm:trt:l}) and (\ref{rm:trt:effect})].  The confidence interval of the restricted mean treatment effect ($\hat{\gamma}(\tau)$) can then be estimated by the $2.5^{th}$ and $97.5^{th}$ percentiles of the bootstrap distribution.

\bigskip
\noindent
{\bf Remark 1.} The Brier Score measures agreement at only one particular time.  As such, the set of values (i.e., $t_1,..., t_s$) over which the Brier Score is minimized [see equation (\ref{bs:stacking:t})] has implications for performance.  In particular, care should be taken to avoid picking only very small, or very large values.   \citet{wey:others:2013} recommend at least nine evenly spaced quantiles of the observed event distribution to ensure good estimation of the conditional survival function.  The effect of the set of time points over which the Brier Score is minimized is investigated in the Supporting Information.

\bigskip
\noindent
{\bf Remark 2.} \citet{wey:others:2013} show that, given the stack contains a uniformly consistent estimator of the conditional survival function, the stacked estimator is uniformly consistent for the underlying conditional survival function.  Therefore, when at least one model within the set of candidate survival models is correctly specified, $\hat{\gamma}(\tau)$ is consistent for the true restricted mean treatment effect by the dominated convergence theorem \citep{ferguson:1996}.  The proposed estimator is, therefore, consistent in a wider range of scenarios than previous methods that assume a proportional hazards model to estimate the conditional survival distribution.

\section{Simulation Study}
\label{rm:sim:study}

\subsection{Set-up}
\label{rm:sim:setup}
The simulation study evaluates the finite sample performance for estimating the causal restricted mean treatment effect with stacked survival models.  We consider four different data-generating scenarios, indexed by $b$, for the covariate-adjusted survival distribution of the potential outcomes.  When $b = 1, 2$ then $T^{a}_b \sim \text{Exp} [ \exp \{ \lambda_b^{a} \} ]$, where $E\{ T^{a}_b \} = 1 / \exp \{ \lambda_b^{a} \}$; when $b = 3, 4$ then $T^{a}_b \sim \text{Gamma}[ \text{scale} = \exp \{ \lambda_b^{a} \}, \text{shape} = 2.5]$, where $E\{ T^{a}_b \} = 2.5 \cdot \exp \{ \lambda_b^{a} \}$.  The covariate effects for the control group (i.e., $\lambda_b^{0}$) are
\begin{eqnarray}
\lambda_1^{0} &=& -4.50 - 0.125 \cdot \{ x_1 + x_2 + x_3 + x_4 \} \nonumber \\  
\lambda_2^{0} &=& -0.70 - 1.000 \cdot \{ \Phi(4 \times x_1) + \Phi(4 \times x_2) + \Phi(4 \times x_3) + \Phi(4 \times x_4) \} \nonumber \\
\lambda_3^{0} &=& ~3.50 - 0.125 \cdot \{ x_1 + x_2 + x_3 + x_4 \} \nonumber \\
\lambda_4^{0} &=& ~4.50 - 0.500 \cdot \{ \Phi(4 \times x_1) + \Phi(4 \times x_2) + \Phi(4 \times x_3) + \Phi(4 \times x_4) \}, \nonumber
\end{eqnarray}
where $\Phi(\cdot)$ is the cumulative distribution function of a standard normal distribution (i.e., the non-linear effect is a `smooth step function').  The covariate effects for the treatment group (i.e., $\lambda_b^{(1)}$) are
\begin{eqnarray}
\lambda_1^{1} &=& -3.50 - 0.125 \cdot \{ x_1 + x_2 + x_3 + x_4 \}  \nonumber \\
\lambda_2^{1} &=& -1.70 - 1.000 \cdot \{ \Phi(4 \times x_1) + \Phi(4 \times x_2) + \Phi(4 \times x_3) + \Phi(4 \times x_4) \} \nonumber \\
\lambda_3^{1} &=& ~3.00 - 0.125 \cdot \{ x_1 + x_2 + x_3 + x_4 \}  \nonumber \\
\lambda_4^{1} &=& ~4.00 - 0.500 \cdot \{ \Phi(4 \times x_1) + \Phi(4 \times x_2) + \Phi(4 \times x_3) + \Phi(4 \times x_4) \}. \nonumber
\end{eqnarray}
The censoring distributions are defined similarly with $\lambda_b^{(a)}$ replaced by $\gamma_b^{(a)}$, and are designed to achieve a marginal censoring rate of approximately $30\%$.  The censoring distributions for the control group (i.e., $\gamma^{(0)}$) are
\begin{eqnarray}
\gamma_1^{0} &=& -4.895 - 0.0625 \cdot \{ x_1 + x_2 + x_3 + x_4 \} \nonumber \\
\gamma_2^{0} &=& -3.680 - 0.5000 \cdot \{ \Phi(4 \times x_1) + \Phi(4 \times x_2) + \Phi(4 \times x_3) + \Phi(4 \times x_4) \} \nonumber \\
\gamma_3^{0} &=& ~3.780 - 0.0625 \cdot \{ x_1 + x_2 + x_3 + x_4 \}   \nonumber \\
\gamma_4^{0} &=& ~4.765 - 0.5000 \cdot \{ \Phi(4 \times x_1) + \Phi(4 \times x_2) + \Phi(4 \times x_3) + \Phi(4 \times x_4) \}, \nonumber
\end{eqnarray}
while the censoring distributions for the treatment group (i.e., $\gamma^{(1)}$) are
\begin{eqnarray}
\gamma_1^{1} &=& -5.395 - 0.0625 \cdot \{ x_1 + x_2 + x_3 + x_4 \} \nonumber \\
\gamma_2^{1} &=& -4.680 - 0.5000 \cdot \{ \Phi(4 \times x_1) + \Phi(4 \times x_2) + \Phi(4 \times x_3) + \Phi(4 \times x_4) \} \nonumber \\
\gamma_3^{1} &=& ~3.280 - 0.0625 \cdot \{ x_1 + x_2 + x_3 + x_4 \}    \nonumber \\
\gamma_4^{1} &=& ~4.265 - 0.5000 \cdot \{ \Phi(4 \times x_1) + \Phi(4 \times x_2) + \Phi(4 \times x_3) + \Phi(4 \times x_4) \}.   \nonumber
\end{eqnarray}
For brevity, we refer to scenarios 1 and 2 as, respectively, the linear and non-linear exponential scenarios, and scenarios 3 and 4 as, respectively, the linear and non-linear gamma scenarios.

The covariate distribution is a four-dimensional multivariate normal with mean zero, unit variances, and a positive $AR(1)$ correlation structure ($\rho = 0.4$).  To mimic observational studies with confounding, the treatment assignment depends on the covariate distribution.  In particular, the probability of receiving treatment, i.e., $P(a_i = 1 | \boldsymbol{x}) = p_i$, is defined by $\text{logit}(p_i) = 0.5 \times (x_1 + x_2 + x_3 + x_4)$.  

Each simulation scenario evaluates performance when $\tau = 20$ and $\tau = 50$ with $1,000$ replications and a sample size of $300$.  The bootstrap distributions are estimated with $300$ bootstrap replicates.  All simulations were run in R version 3.0.0 \citep{R}.  The stacking weights are estimated by minimizing the Brier Score over nine equally spaced quantiles of the observed events with the constrained minimization problem solved by the \texttt{alabama} package \citep{alabama:2013}.

We consider a mixture of parametric, semi-parametric and non-parametric candidate survival models.  The parametric models are the Weibull model and log-Normal model with only linear main effects.  Both models are special cases of an accelerated failure time model, while the Weibull is also a special case of a proportional hazards model.  The semi-parametric models are two versions of the Cox model.  The first Cox model has only linear main effects, while the second Cox model uses penalized splines for main effects with the roughness penalty set to $0.5$.  The \texttt{survival} package estimates both the parametric and semi-parametric models \citep{survival}.  The non-parametric estimator in the set of candidate survival models is random survival forests (RSF), which is estimated with the \texttt{randomSurvivalForest} package \citep{ishwaran:others:2008} with $1,000$ trees grown.   We use the default tuning parameters of the \texttt{randomSurvivalForest} package for RSF.  

We consider two different versions of stacked survival models.  The first version excludes random survival forests (RSF) from the set of candidate survival models, while the second includes RSF.  In particular, 
\begin{enumerate}
	\item The `Stacked' estimator only includes the two parametric and two semi-parametric survival models in the set of candidate survival models.
	\item The `Stacked (with RSF)' estimator includes the two parametric models, two semi-parametric models, and one non-parametric  model with default tuning parameters.
\end{enumerate}
There are two significant motivations for investigating two different versions of the stacked estimator.  The first motivation is the potential reduction of computational demands.  In particular, random survival forests (RSF) are computationally expensive, which is significantly compounded by the non-parametric bootstrap used to estimate confidence intervals.  We can save substantial computational time if the stacked estimator without RSF performs as well, or better, than the stacked estimators with RSF.  The second motivation is that sample size of $300$ is potentially too small for a non-parametric estimator, such as RSF, to perform well enough to contribute to a better stacked estimator.

Each restricted mean estimator in this simulation study uses the ``regression'' approach described in Section \ref{sub:sec:rmte}.  The different methods of estimating the restricted mean differ in their approach to estimating the treatment-specific conditional survival functions in equation (\ref{rm:trt:l}).  The two versions of the stacked estimator are compared to a Cox proportional hazards model with linear main effects (referred to as the `Cox estimator'), and a Cox proportional hazards model with penalized splines (referred to as the `Splines estimator').  The Cox estimator was proposed by \citet{chen:tsiatis:2001}, while the Splines estimator is a straightforward extension of the Cox estimator that should be more robust in a variety of situations.  Note that each stacked estimator includes both the Cox and Splines estimators in the set of candidate survival models.

The methods are compared on the basis of percent relative bias, i.e., $100 \times [E\hat{\delta}(\tau) - \delta(\tau)] / \delta(\tau)$, and the ratio of mean squared error (MSE) to the Cox estimator, where $\text{MSE} = E\{\hat{\delta}(\tau) - \delta(\tau)\}^2$.  Confidence interval performance is assessed with two measures: the ratio of average confidence interval length (ACL) to the Cox estimator and coverage probability.  For each method, the estimated confidence intervals use the $2.5^{th}$ and $97.5^{th}$ percentiles of the bootstrap distribution with $300$ bootstrap replicates.  We also present the ratio of `integrated squared survival error' (ISSE) to the Cox estimator for each method: $\text{ISSE}_{\tau}\{ \hat{S}(\cdot | \boldsymbol{x}) \} = \frac{1}{2} \times E \int_0^{\tau} \{ [ \hat{S}^{(0)}(t | \boldsymbol{x}) - S^{(0)}(t | \boldsymbol{x}) ]^2 + [ \hat{S}^{(1)}(t | \boldsymbol{x}) - S^{(1)}(t | \boldsymbol{x}) ]^2 \} dt$, which corresponds to the average of the mean-squared error of the treatment-specific conditional survival functions presented in Section~\ref{cdf:influence}.  All expectations are approximated by averaging across the $1,000$ replications.

In the exponential scenario with linear covariate effects, the correctly specified Cox estimator should perform well, while the Splines estimator - despite being correctly specified - should be slightly less efficient due to increased flexibility. In contrast, for the exponential scenario with non-linear covariate effects, the Cox estimator should perform poorly due to model misspecification, while the Splines estimator remains correctly specified and should perform relatively well.  Despite including both the Cox estimator and the Splines estimator, both stacked estimators will likely perform relatively worse in both exponential scenarios due to the increased flexibility of additional models in the stack.  

Each parametric and semi-parametric model is misspecified in both gamma scenarios.  However, the parametric models in the stacked estimators closely approximate the truth in the linear scenario (e.g., same mean function).  As such, in the linear gamma scenario, the stacked estimators should perform better than both the Cox and Splines estimator due to approximately correct parametric models.  The non-linear gamma scenario assesses the robustness of each estimator since none of the parametric and semi-parametric models are correctly specified or closely approximate the underlying mean function.  However, the stacked estimators should perform better than the Cox and Splines estimators due to the increased flexibility of additional models in the set of candidate models.  This is also an opportunity for the stacked estimator with random survival forests (RSF) to perform well relative to the Stacked estimator without RSF.

\subsection{Results}
\label{rm:sim:results}
Across the different scenarios, the Stacked estimator (without RSF) possesses as good, or better, bias and MSE than the Stacked estimator with RSF.  The Stacked estimator with RSF also, at times, does not achieve nominal coverage at this modest sample size.  In addition, the Stacked estimator with RSF actually performs slightly worse than the Stacked estimator (without RSF) in the non-linear gamma scenario; an advantageous scenario for random survival forests (RSF).  Due to the similar, or better, performance, we only compare the Cox and Splines estimators to the Stacked estimator (without RSF) for the rest of this section.

For the exponential scenarios (Table \ref{tab:rm:var4:expo}), the Stacked estimator possesses similar, or slightly more, bias than the Cox estimator when the covariate effects are linear, and the Stacked estimator has similar, or more, bias than the Splines estimator for both linear and non-linear covariate effect scenarios.  These results are expected as the Cox and Splines estimators are correctly specified in, respectively, the linear and non-linear exponential scenarios.  In the exponential scenario with non-linear covariate effects, the Stacked estimator has approximately $\sim 4\%$ lower relative bias than the misspecified Cox estimator.    In the linear scenario, the Stacked estimator has similar, or slightly larger, MSE than the Cox estimator but, in the non-linear scenario, the Stacked estimator has approximately $20\%$ lower MSE than the Cox estimator.  In addition, the Stacked estimator possesses approximately $10\%$ lower MSE than the Splines estimator in both exponential scenarios.  Finally, the Stacked estimator possesses approximately $5\%$ narrower confidence intervals than the Splines estimator in both scenarios including, surprisingly, the non-linear scenario.  The Stacked estimator is therefore competitive in the exponential scenarios with linear effects and more efficient than both the Cox and Splines estimators in the non-linear exponential scenario.  

For both gamma scenarios (Table \ref{tab:rm:var4:gamma}), the Stacked estimator possesses $5\%-12\%$ less relative bias than the Cox estimator and $1\%-6\%$ less relative bias than the Splines estimator.  In addition, the Stacked estimator possesses $8\%-20\%$ lower MSE than the Cox estimator and $20\%-30\%$ lower MSE than the Splines estimator.  The Cox estimator surprisingly possesses lower MSE than the Splines estimator and as good, or better, ACL than the Stacked estimator in almost every gamma scenario.  However, aside from ACL, the Stacked estimator performs better than both the Cox and Splines estimators under non-proportional hazards.

\citet{wey:others:2013} motivated stacking with the goal of balancing, for a given sample size, the low variance of (potentially misspecified) parametric models with robust (but potentially inefficient) semi-parametric and non-parametric models.  This effect is illustrated in the simulation study even when the random survival forests (i.e., the non-parametric estimator) are excluded from the set of candidate survival models.  For example, in the linear exponential scenario, the Stacked estimator possesses as good, or better, MSE than the correctly specified Cox and Splines estimators.  This is likely due to the inclusion of a correctly specified parametric Weibull model.  Yet, even the Weibull model is misspecified in the non-linear gamma scenario, the Stacked estimator still performs better than both the Cox and Splines estimators despite the lack of either a correctly specified parametric model or a non-parametric estimator.  This ability to adaptively find a good balance between the low variance of (potentially misspecified) parametric models with the more robust, but still potentially misspecified, semi-parametric models (e.g., a Cox model with penalized splines) is the most appealing aspect of stacked survival models.

\begin{figure}[hptb]
\centering
\caption{An investigation into the relationship between restricted mean treatment effect performance and the quality of the conditional survival function estimation, which is measured by integrated squared survival error, or $\text{ISSE} = E E_{\boldsymbol{x}} \{ \int_0^{\tau} [ \hat{S}(t | \boldsymbol{x}) - S(t | \boldsymbol{x}) ] ^ 2 dt \}$.}
\includegraphics[height=5cm,width=5cm]{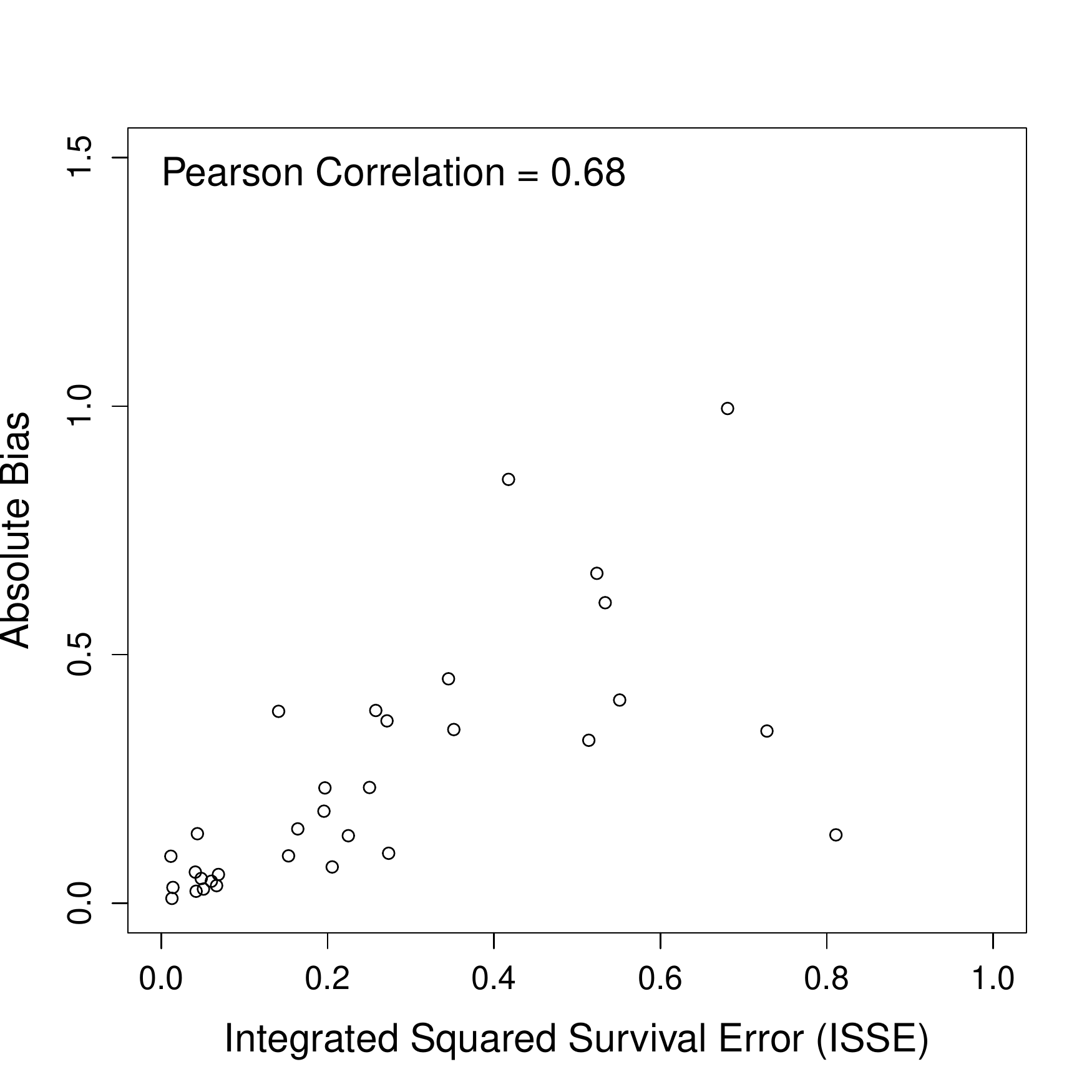}
\includegraphics[height=5cm,width=5cm]{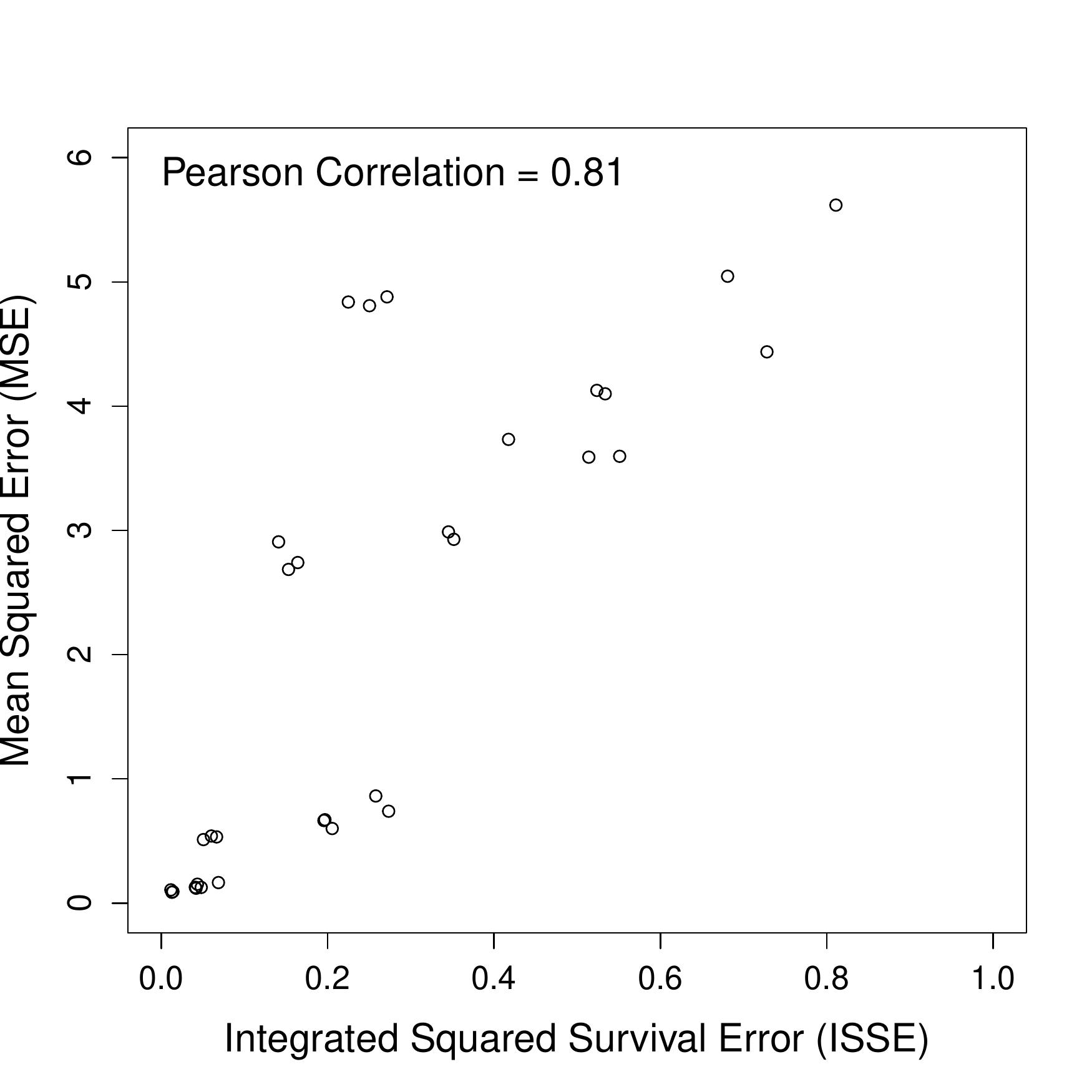}
\label{fig:rm:perf:isse}
\end{figure}

Section~\ref{cdf:influence} argues that performance of the treatment-specific conditional survival functions will be more tightly associated with MSE, rather than bias, of the restricted mean treatment effect.  Figure~\ref{fig:rm:perf:isse} provides an illustration of this point from the simulation study.  In particular, the Pearson correlation of the ISSE with the MSE of the restricted mean treatment effect is $0.81$, while the correlation of the ISSE with the bias of the restricted mean treatment effect is $0.68$.  As illustrated in the Supporting Information, the correlation at a larger sample size increases for the MSE, while correlation slightly decreases for the bias.

\noindent
{\bf Remark 3.} The Supporting Information contains several additional investigations into the simulation study: a larger sample size, the difference between the bootstrap standard error and Monte Carlo standard error, different approaches to confidence interval estimation, and the choice of time points over which the stacking weights are determined.

\section{Effect of Center Volume in Lung Transplantation}
\label{rm:LT:applied}
We applied the proposed estimator to data from an observational registry of post-lung transplant survival to estimate the effect of large center volume on graft survival (i.e., time to death or retransplantation).  In particular, we want to estimate the difference in the restricted mean of post-transplant survival between high volume centers (defined as $> 100$ lung transplants over the past two years \citep{tsuang:others:2013}) and low volume centers (defined as $\leq 100$ lung transplants over the past two years).  However, previous research has demonstrated that the relationship between transplant type, which is an important confounder, and the post-transplant hazard is likely non-proportional \citep{thabut:others:2009}.  Therefore, this example represents a setting for estimating restricted mean treatment effects with stacked survival models instead of the Cox proportional hazards model.

The United Network for Organ Sharing (UNOS) collects patient information, donor information and survival status of every solid organ transplant performed in the United States.  This analysis only includes lung transplants performed between January 1, 2008 and December 31, 2011 in adult recipients receiving their first lung-only transplantation.  We adjust for potential confounding from several patient related covariates including gender, age, lung allocation score, native disease grouping (obstructive, vascular, cystic and restrictive), distance walked in six minutes, ventilator use, level of oxygen use, and type of transplant (single versus bilateral lung transplant).  We also adjust for several donor related covariates: age over 55 years, African American race, smoking history greater than $20$ pack years, and height difference between donor and recipient.  The event of interest is time to death or retransplantation.  A total of $5,499$ transplanted patients were included in this analysis.  Approximately $76\%$ of the survival times were censored.  

Similar to Stacked estimator (without random survival forests) in Section \ref{rm:sim:study}, the stacked survival model includes two versions of the Cox model, a Weibull model, and a log-Normal model.  The Weibull, log-Normal, and the first Cox model fit linear main effects to all continuous covariates except the height difference between donor and recipient, which is fit with a quadratic main effect.  The second Cox model fits penalized splines to each continuous covariate.  The $95\%$ confidence intervals are estimated with $2.5^{th}$ and $97.5^{th}$ quantiles of the bootstrap distribution with $1000$ bootstrap replications.

\begin{figure}[hptb]
\centering
\caption{The estimated difference in restricted means between high volume centers and low volume centers across a range of $\tau$ vaues.  The top graph plots the estimated restricted mean treatment effects for the Cox and Stacked estimators, while the bottom graph standardizes the restricted mean treatment effect by $\tau$ (i.e., $\hat{\gamma}(\tau) / \tau$).  The dashed lines are the $95\%$ confidence interval limits.  The non-parametric bootstrap estimates the confidence intervals with the $2.5^{th}$ and $97.5^{th}$ quantiles of the bootstrap distribution.  The Stacked estimator does {\it not} include random survival forests.}
\includegraphics[height=5cm,width=5cm]{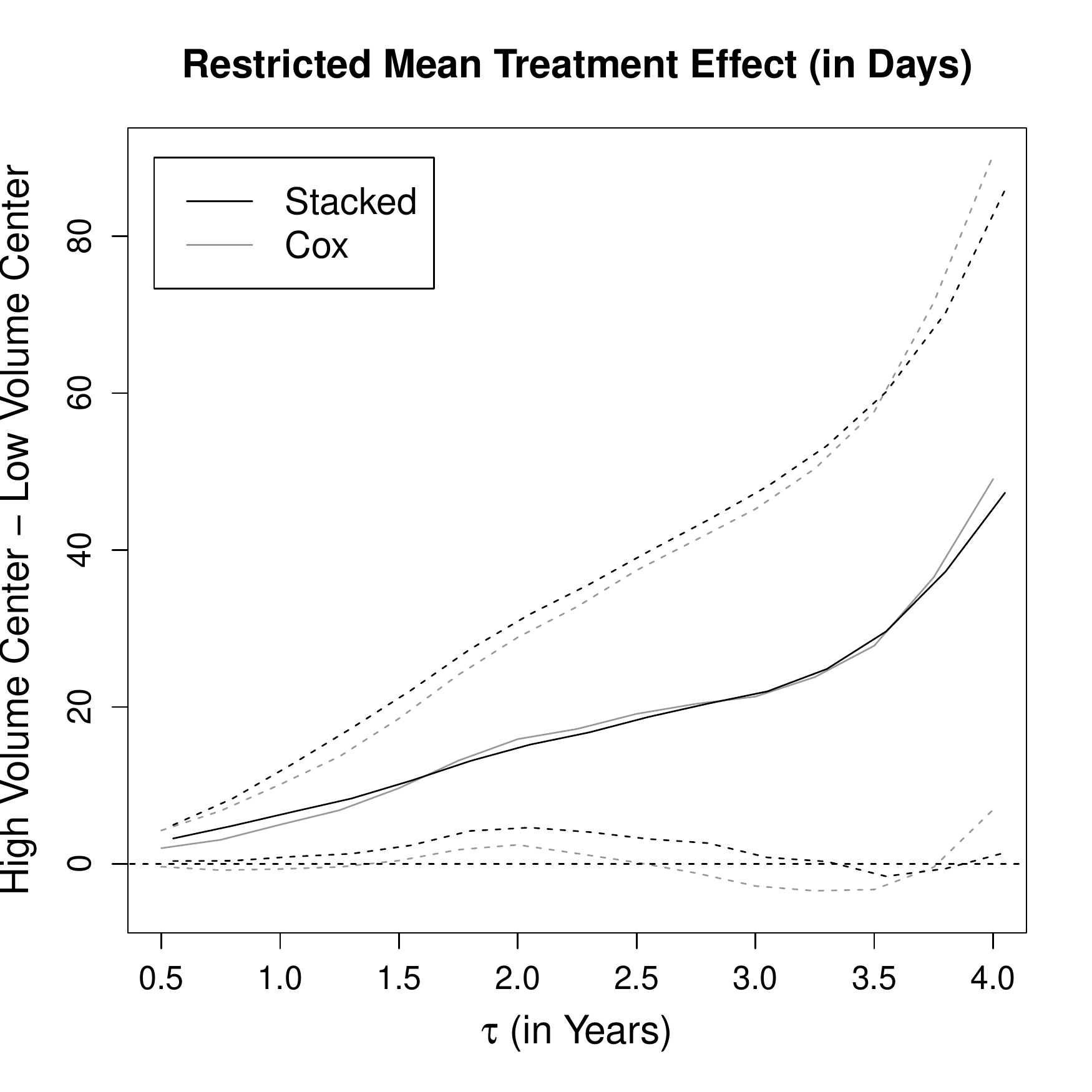}
\includegraphics[height=5cm,width=5cm]{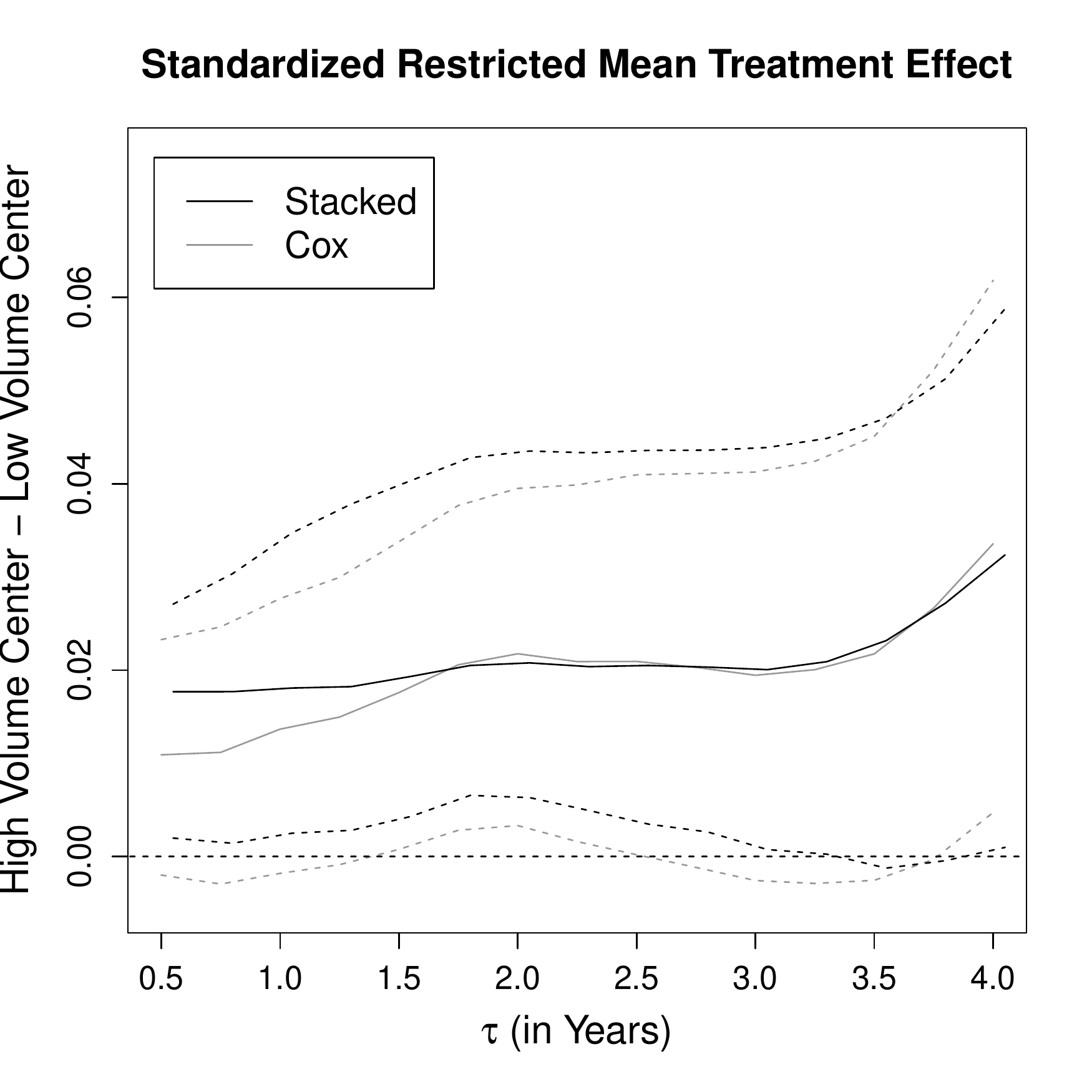}
\label{fig:gbcs:rms}
\end{figure}

Figure \ref{fig:gbcs:rms} gives the estimated causal restricted mean treatment effect for high volume versus low volume centers from $\tau = 0.5$ years to $\tau = 4$ years based on the Stacked estimator and the Cox estimator.  Both the Stacked and Cox estimators estimate a restricted mean difference between large and small volume centers greater than zero over the range of follow-up, which indicates better post-transplant survival for high volume centers.  The Stacked estimate is approximately $30-60\%$ larger than the Cox estimate from $0.5$ years to $1$ years, while at $4$ years the Stacked estimate is approximately $5\%$ smaller than the Cox estimate.  In addition, the confidence intervals for the Stacked estimator do not include zero up to about $3$ years, while the confidence intervals for the Cox estimator contain zero until about $1.5$ years and becomes non-significant only from $2.5$ to $3.5$ years.  As illustrated in the simulations, the difference in significance between the two estimators may be a function of the consistently lower MSE of the Stacked estimate in non-proportional hazards scenarios.

\citet{weiss:others:2009} were specifically interested in the mortality risk for center volume at one year.  Since earlier survival is indicative of peri-operative and early post-operative mortality (which is relatively high for lung transplantation), the one-year restricted mean survival is a clinically important outcome.  The Stacked estimator suggests a $30\%$ larger difference than the Cox estimator in the causal restricted mean survival over one year.  In addition, the confidence interval for the Cox estimate includes zero, while that the Stacked estimator does not.

\section{Concluding Remarks}
\label{rm:conclusion:sec}  
We frame the estimation of causal restricted mean treatment effects as a problem of estimating the conditional survival function.  In most application areas, there is little information to suggest {\it a priori} an appropriate distributional assumption for the survival time or functional form for the covariates.  This motivates flexibly estimating restricted mean treatment effects for observational studies with stacked survival models.  In particular, we demonstrate that stacked survival models successfully estimates an optimal combination of candidate survival models that performs well at a given sample size.

Restricted mean survival is traditionally estimated by first estimating the conditional survival distribution with a Cox proportional hazards model, yet the simulation study illustrates that there is little cost to considering the more flexible stacked survival models given that they achieve similar mean squared error (MSE) under proportional hazards.  When the proportional hazards assumption is violated, stacked survival models can substantially reduce the bias and MSE of the causal restricted mean treatment effect.  In addition, the stacked estimator is consistent in a wider range of situations compared than current approaches that assume a proportional hazards model.  We also demonstrated that the Stacked estimator identifies a clinically meaningful difference in lung transplantation between high-volume and low-volume centers for the one year restricted mean, while the proportional hazards estimator fails detect the difference.

We considered two different approaches to stacked survival models: one estimator (the `Stacked' estimator) excluded the non-parametric random survival forests (RSF) from the set of candidate models, and the other estimator included RSF in the set of candidate models.  We found that the Stacked estimator (without RSF) performed as good, or better, and was computationally faster than the stacked estimator with RSF; these relationships held even when the sample size was increased to $n = 900$ (see the Supporting Information).  Thus, the Stacked estimator without RSF was a reasonable choice for estimating restricted means with stacked survival models in the simulation scenarios considered here.  However, different scenarios may exhibit meaningful differences in performance when RSF or other non-parametric estimators are included in the set of candidate survival models.  This point illustrates that the appropriate selection of candidate survival models remains a significant area of future research for stacked survival models.  

There are two main approaches to estimating the causal restricted mean treatment effect: the ``regression approach'' pursued in this paper and an approach based on inverse-probability weighting (IPW) for treatment assignment and censoring \citep{schaubel:wei:2011,zhang:schaubel:2012}.  The IPW approach requires forming models for the censoring and treatment distributions.  The ``regression'' approach is a more efficient estimator of the restricted mean difference when the conditional survival model has been correctly specified.  However, the IPW approach is sometimes preferred because standard methods that estimate the conditional survival distribution may be overly restrictive (e.g., the Cox proportional hazards model).  The flexibility of stacked survival models may mitigate some of the concerns of the ``regression'' approach.  


Statisticians regularly want to form linear contrasts of the restricted mean survival.  A common approach to restricted mean regression uses pseudo-observations from the leave-one-out jackknife of the Kaplan-Meier restricted mean estimator as the outcome variable in a generalized linear model \citep{andersen:others:2003, andersen:others:2004}.  The potential benefit for estimating pseudo-observations with stacked survival models is not clear.  However, an alternative to pseudo-values is the model-free contrast approach proposed by \citet{rudser:leblanc:emerson:2012}, which would likely benefit from stacked survival models.  The investigation into these areas is a future research interest.

There has been recent research on using model averaging to account for uncertainty in the confounders of the treatment-outcome relationship \citep{wang:others:2012, cefalu:others:2013}.  However, previous work has assumed that the structure of the relationship between the covariates and outcome was known (e.g., linear relationship between covariates and log-hazard) although, in practice, there is usually little evidence to support {\it a priori} assumptions on the survival outcomes and functional form of the covariate.  We demonstrate that principally averaging different model structures can lead to substantially better performance in the estimation of the causal restricted mean treatment effect.  Thus, an interesting avenue for future research would consider the selection of covariates based on both the outcome and treatment models, while also considering different distributional assumptions and functional forms for the outcome model.

A conditional survival function is required by many methods besides restricted mean treatment effects:  for example, censored quantile regression \citep{wey:wang:rudser:2014}, time-dependent ROC curves \citep{zheng:heagerty:2004}, inverse probability-of-censoring weighted estimators \citep{fine:gray:1999}, model-free contrast approaches \citep{rudser:leblanc:emerson:2012}, and dynamic treatment regime methods \citep{zhao:others:2011}.  Similar to restricted mean treatment effect estimation, all of these methods have traditionally used a Cox model or a non-parametric method to estimate the conditional survival function.  The success of stacked survival models for restricted mean treatment effect estimation illustrates the potential improvement for a wide spectrum of survival analysis methods.

\bibliographystyle{biom}
\bibliography{paperbib}

\pagebreak

\begin{table}[hptb]
\footnotesize
\caption{Simulation results for the exponential distributed scenarios: $N = 300$, $N_{SIM} = 1000$, and a marginal censoring of $30\%$.  The confidence intervals are estimated using the $2.5^{th}$ and $97.5^{th}$ quantiles of the non-parametric bootstrap distribution with $300$ bootstrap replicates.  `Percent Relative Bias' is the ratio of the bias and true restricted mean difference.  `MSE Ratio' is the ratio of MSE relative to the Cox estimator.  `ACL Ratio' is the ratio of average confidence interval lengths relative to the Cox estimator.  `Cov' is the coverage probability for the confidence interval.  `ISSE Ratio' is the ratio of integrated squared survival error, which corresponds to the mean-squared of the conditional survival function, relative to the Cox estimator.  The $\gamma(20)$ and $\gamma(50)$ are the true restricted mean treatment effects for $\tau = 20$ and $\tau = 50$, respectively.}
\begin{center}
\begin{tabular}{l l ccccc}
& & Percent & & & &  \\ 
 & Estimator & Relative Bias & MSE Ratio & ACL Ratio & Cov. & ISSE Ratio  \\ \hline

\multirow{4}{3cm}{ Linear \\ $\gamma(20) = -2.965$}
		& Cox				& ~~1$\%$ & 1.00 & 1.00 & 0.95 & 1.00   \\
		& Splines				& ~-2$\%$  & 1.17 & 1.12 & 0.96 & 4.03   \\
		& Stacked				& ~-1$\%$  & 1.06 & 1.05 & 0.96 & 1.19   \\ 
		& Stacked (with RSF)	& ~-1$\%$  & 1.04 & 0.96 & 0.93 & 1.32   \\ \hline

\multirow{4}{3cm}{ Non-Linear \\ $\gamma(20) = 2.690$}
		& Cox				& ~10$\%$ & 1.00 & 1.00 & 0.94 & 1.00   \\
		& Splines				& ~~3$\%$ & 0.86 & 1.04 & 0.95 & 1.06   \\
		& Stacked				& ~~5$\%$ & 0.77 & 0.98 & 0.95 & 0.76   \\ 
		& Stacked (with RSF)	& ~~6$\%$ & 0.78 & 0.86 & 0.88 & 0.76   \\ \hline

& & & & & &  \\  \hline

\multirow{4}{3cm}{ Linear \\ $\gamma(50) = -12.318$}
		& Cox				& ~~1$\%$ & 1.00 & 1.00 & 0.95 & 1.00   \\
		& Splines				& ~~1$\%$ & 1.16 & 1.10 & 0.95 & 3.60   \\
		& Stacked				& ~~2$\%$ & 0.99 & 1.02 & 0.95 & 1.11   \\ 
		& Stacked (with RSF)	& ~~3$\%$ & 1.01 & 0.90 & 0.84 & 1.20   \\ \hline

\multirow{4}{3cm}{ Non-Linear \\ $\gamma(50) = 7.929$}
		& Cox				& ~10$\%$ & 1.00 & 1.00 & 0.92 & 1.00   \\
		& Splines				& ~~3$\%$ & 0.88 & 1.05 & 0.95 & 1.07   \\
		& Stacked				& ~~6$\%$ & 0.82 & 0.98 & 0.94 & 0.77   \\ 
		& Stacked (with RSF)	& ~~6$\%$ & 0.81 & 0.82 & 0.92 & 0.78   \\ \hline

\end{tabular}
\label{tab:rm:var4:expo}
\end{center}
\end{table}

\begin{table}[hptb]
\footnotesize
\caption{Simulation results for the gamma distributed scenarios: $N = 300$, $N_{SIM} = 1000$, and a marginal censoring of $30\%$.  The confidence intervals are estimated using the $2.5^{th}$ and $97.5^{th}$ quantiles of the non-parametric bootstrap distribution with $300$ bootstrap replicates.  `Percent Relative Bias' is the ratio of the bias and true restricted mean difference.  `MSE Ratio' is the ratio of MSE relative to the Cox estimator.  `ACL Ratio' is the ratio of average confidence interval lengths relative to the Cox estimator.  `Cov' is the coverage probability for the confidence interval.  `ISSE Ratio' is the ratio of integrated squared survival error, which corresponds to the mean-squared of the conditional survival function, relative to the Cox estimator.  The $\gamma(20)$ and $\gamma(50)$ are the true restricted mean treatment effects for $\tau = 20$ and $\tau = 50$, respectively.}
\begin{center}
\begin{tabular}{l l ccccc}
& & Percent & & & &  \\ 
 & Estimator & Relative Bias & MSE Ratio & ACL Ratio & Cov. & ISSE Ratio  \\ \hline

\multirow{4}{3cm}{ Linear \\ $\gamma(20) = -0.753$}
		& Cox				& -13$\%$ & 1.00 & 1.00 & 0.93 & 1.00   \\
		& Splines				& ~-7$\%$ & 1.19 & 1.19 & 0.95 & 4.14   \\
		& Stacked				& ~-1$\%$ & 0.82 & 1.05 & 0.95 & 1.11   \\ 
		& Stacked (with RSF)	& ~-4$\%$ & 0.87 & 1.01 & 0.89 & 1.21   \\ \hline

\multirow{4}{3cm}{ Non-Linear \\ $\gamma(20) = -0.931$}
		& Cox				& -15$\%$ & 1.00 & 1.00 & 0.92 & 1.00   \\
		& Splines				& ~-6$\%$ & 1.09 & 1.15 & 0.94 & 1.58   \\
		& Stacked				& ~-3$\%$ & 0.79 & 1.04 & 0.95 & 0.96   \\ 
		& Stacked (with RSF)	& ~-7$\%$ & 0.84 & 0.99 & 0.88 & 0.94   \\ \hline

& & & & & &  \\  \hline

\multirow{4}{3cm}{ Linear \\ $\gamma(50) = -6.599$}
		& Cox				& ~-6$\%$ & 1.00 & 1.00 & 0.93 & 1.00   \\
		& Splines				& ~-5$\%$ & 1.23 & 1.13 & 0.94 & 3.64   \\
		& Stacked				& ~-1$\%$ & 0.92 & 1.04 & 0.94 & 1.08   \\ 
		& Stacked (with RSF)	& ~-2$\%$ & 0.94 & 0.96 & 0.87 & 1.16   \\ \hline

\multirow{4}{3cm}{ Non-Linear \\ $\gamma(50) = -6.407$}
		& Cox				& -13$\%$ & 1.00 & 1.00 & 0.91 & 1.00   \\
		& Splines				& ~-6$\%$ & 0.96 & 1.07 & 0.94 & 1.32   \\
		& Stacked				& ~-5$\%$ & 0.78 & 1.00 & 0.94 & 0.84   \\ 
		& Stacked (with RSF)	& ~-7$\%$ & 0.80 & 0.91 & 0.86 & 0.83   \\ \hline

\end{tabular}
\label{tab:rm:var4:gamma}
\end{center}
\end{table}

\end{document}